\documentclass[prd,aps,nofootinbib]{revtex4}
\usepackage{graphicx,subfigure}
\usepackage{amsmath,amssymb}

\def\beq{\begin{equation}}
\def\eeq{\end{equation}}
\def\beqr{\begin{eqnarray}}
\def\eeqr{\end{eqnarray}}
\def\bdpm{\begin{displaymath}}
\def\edpm{\end{displaymath}}

\newcommand{\aem}{\alpha}

\newcommand{\nnb}{\nonumber}

\begin{document}

\title{\Large \bf Right-handed current contributions in $B \to K \pi$ decays}
\date{\today}

\author{ Kihyeon Cho }
\email{cho@kisti.re.kr}

\author{ Soo-hyeon Nam }
\email{glvnsh@gmail.com}

\affiliation{
National Institute of Supercomputing and Networking, KISTI, Daejeon 305-806, Korea
}

\date{\today}

\begin{abstract}
We reexamine the right-handed current effects in $b \to s$ transitions in nonmanifest left-right models.
Using the effective Hamiltonian approach including all possible low-energy operators, 
we obtain especially the $B \to K \pi$ decay amplitudes including annihilation contributions,
and investigate the right-handed current contributions to CP asymmetries in $B \to K \pi$ decays.
Taking into account the constraints from global analysis of muon decay measurements,
$|V_{ub}|$ measurements in inclusive and exclusive B decays, and $B_s^0-\bar{B_s^0}$ mixing measurements,
we find the allowed regions of new physics parameters satisfying the current experimental data.
\end{abstract}
\maketitle

\section{Introduction}

  CP asymmetry measurements in nonleptonic $b \to s$ decays have been receiving considerable attention
over the past several years 
since the recent experimental measurements in some decay channels are in disagreement 
with naive estimates of the Standard Model (SM).
One of the important examples is the direct CP asymmetries in $B \to K \pi$ decays \cite{Mishima11}.
Current world averages of the CP asymmetries in $B \to K \pi$ decays are given by \cite{HFAG}:
\beqr \label{ACP_exp}
A_{CP}(B^0 \to K^\pm \pi^\mp) &=& -0.086 \pm 0.007 , \cr
A_{CP}(B^\pm \to K^\pm \pi^0) &=& 0.040 \pm 0.021 ,\cr
A_{CP}(B^\pm \to K^0 \pi^\pm) &=& -0.015 \pm 0.012 .
\eeqr
However, the naive factorization assumption predicts
$A_{CP}(B^0 \to K^\pm \pi^\mp) \approx A_{CP}(B^\pm \to K^\pm \pi^0)$ \cite{Ali98},
which is inconsistent with the current data in Eq. (\ref{ACP_exp}).
This discrepancy can be explained by enhancing the smaller diagrams such as $C'$ and $P'_{EW}$ 
with a sizable strong phase through the SM fit to the $K\pi$ data \cite{Baek09},
where $C'$ and $P'_{EW}$ stand for the color-suppressed tree and electroweak penguin amplitudes, respectively,
in the topological decomposition \cite{Gronau94}.   
Nonetheless, such enhancement of subdominant diagrams in the SM is not fully understood theoretically 
and also may not be sufficient to resolve other puzzles simultaneously in the $B$ meson system \cite{Mishima11}.
Alternatively to the SM fit,
since the prediction given in Ref. \cite{Ali98} did not incorporate all possible hadronic uncertainties,
these decay modes have been also studied within the SM in the framework of different factorization approaches
such as QCD factorization \cite{QCDF}, perturbative QCD (PQCD) \cite{PQCD}, and soft-collinear effective theory \cite{SCET}.
Even under such factorization assumptions, however, the above data have not been fully explained as well.
In the SM, the sizes and patterns of CP violation in various decay modes are governed by
a single complex phase which resides in the Cabibbo-Kobayashi-Maskawa (CKM) matrix,
but such large CP violation effects have not been simply explained with this single parameter
in any of those factorization methods in various decay modes simultaneously.
Therefore, there has been several efforts to understand such large CP asymmetries beyond the SM
with additional CP odd parameters \cite{ACP_NP}.  
Similarly in this paper, we study the new physics (NP) contributions to the direct CP asymmetries in $B \to K \pi$ decays 
as well as to (semi-)leptonic $B$ decays and $B$ mixing where NP effects could be sizable.
In order to minimize hadronic uncertainties, 
we consider all relevant tree and penguin contributions even including annihilation types
by adopting PQCD approach, and estimate the possible NP contributions.

  One of the simplest extensions of the SM corresponding to such a scenario with additional CP phases is
the nonmanifest ($V^R \neq V^L$) left-right model (LRM) with gauge group
$SU(2)_L \times SU(2)_R \times U(1)$ where $V^L(V^R)$ is the left(right)-handed quark mixing matrix \cite{LRM}.
Since the LRM has the extended group $SU(2)_R$, there are new parameters such as
a right-handed gauge coupling $g_R$, new charged (neutral) gauge bosons $W_R$ ($Z_R$),
and the $W_L-W_R$ ($Z_L-Z_R$) mixing angle $\xi$ ($\eta$).
After spontaneous symmetry breaking, the gauge eigenstates $W_R$ mix with $W_L$ to form
the mass eigenstates $W$ and $W^\prime$ with masses $M_W$ and $M_{W^\prime}$, respectively.
Similarly, the neutral gauge bosons mix each other \cite{Chay99}, but we do not present them here
because $Z_R$ contribution to flavor-changing B decays is negligible.
Although tree-level flavor-changing neutral Higgs bosons with masses $M_H$ enter into our theory
due to gauge invariance, we also neglect their contributions by assuming $M_H \gg M_{W^\prime}$ \cite{Chang84}.
The mixing angle $\xi$ and the ratio $\zeta$ of $M_W^2$ to $M_{W^\prime}^2$ are
restricted by a number of low-energy phenomenological constraints \cite{Langacker89}.
One of the most stringent bounds on $M_{W'}$ was obtained from $K_L - K_S$ mixing.
If the model has manifest ($V^R = V^L$) left-right symmetry ($g_R = g_L$)
where $V^L(V^R)$ is the left(right)-handed quark mixing matrix, $M_{W'} > 2.5$ TeV \cite{Zhang07}.
Similar bounds were obtained recently by CMS and ATLAS from direct searches for the decay channels of
the extra gauge bosons $W' \rightarrow \ell\nu$
under various assumptions on the right-handed neutrino masses and gauge couplings \cite{LHCW}.
However, the form of $V^R$ is not necessarily restricted to manifest or pseudomanifest ($V^R = V^{L*}K$) symmetric-type,
where $K$ is a diagonal phase matrix \cite{LRM}.
If $V^R$ takes one of the following forms, the $W_R$ mass limit can
be significantly lowered \cite{Langacker89}, and
$V^R_{ub}$ can be as large as $\lambda$ (for $M_{W_R}\geq 800$ GeV)
\cite{Rizzo98}:
  \beq \label{VR3}
\left( \begin{array}{ccc}   1 &  0 &  0 \\
                              0 &   0 &   1 \\
                         0 &  1 &  0 \end{array} \right) ,\quad
\left( \begin{array}{ccc}  0 &  1 &  0 \\
                             1 &  0 &  0 \\
                       0 &  0 & 1 \end{array} \right) ,\quad
\left( \begin{array}{ccc}  0 &  1 &  0 \\
                             0 & 0 &  1 \\
                        1 &  0 &  0 \end{array} \right) \,.
 \eeq

 As well as the CP-violating observables in $B \to K \pi$ decays,
we also accommodate a large CP-violating phase in $B_s$ mixing observed at Tevatron
and a disagreement emerged between the determination of $|V_{ub}|$ from inclusive and exclusive B decays  \cite{Nam08}.
In order to incorporate all of those considerations, we take the following form of $V^R$ as similarly done in Ref. \cite{Buras11}:
\beq
V^R = \left( \begin{array}{ccc}  \sim 0 & c_R e^{i\alpha_1} & s_R e^{i\alpha_2} \\
                            e^{i\omega} & \sim 0  & \sim 0 \\
                       \sim 0 & -s_R e^{i\alpha_3} & c_R e^{i\alpha_4} \end{array} \right),
\label{VR}
\eeq
where $c_R\ (s_R)\equiv \cos\theta_R\ (\sin\theta_R)$ $(0^\circ \leq \theta_R \leq 90^\circ )$.
Here the matrix elements indicated as $\sim 0$ may be $\lesssim 10^{-2}$ and unitarity
requires $\alpha_1+\alpha_4=\alpha_2+\alpha_3$.
Especially, with this form, the present experimental measurement of the large branching fractions for $B \to \tau \nu$ decays can be explained \cite{Nam08,Buras11},
but the right-handed current effect in $B_d$ mixing is negligible so we only consider direct CP asymmetries 
in $B \to K \pi$ decays.
One can of course take different types of $V^R$ without taking into account of $|V^R_{ub}|$,
and relevant studies were done earlier in Refs. \cite{Nam03, Lee12}.

This paper is organized as follows.
In Sec. II, we briefly discuss some of phenomenological constraints without assuming manifest (or pseudomanifest) left-right symmetry.
We present the effective Hamiltonian describing $\Delta B = 1$ and $\Delta S = 1$ transition in Sec. III,
and obtain $B \to K \pi$ decay amplitudes including all relevant tree and penguin contributions 
in the general LRM in Sec. IV.
In Sec. V, we explicitly show the allowed regions of NP parameters satisfying the current experimental data, 
taking into account all the constraints obtained in Sec. II. 
Finally, we conclude in Sec. VI.

\section{Phenomenological Constraints}

  We consider the case that $W'$ masses are not too heavy so that it can be accessible at LHC.
Without assuming manifest or pseudomanifest left-right symmetry in the general LRM,
$W'$ masses are not highly constrained by low-energy electroweak measurements,
but still $W'$ exchange effects could be seen in various decay modes,
and the bound of its mass could be obtained independent of the form of $V^R$.
For instance, we can obtain the lower bound on $M_{W^\prime}$ from global analysis of muon decay measurements
as follows \cite{Hillairet12}:
\beq
\zeta_g < 0.017 \qquad \textrm{or} \qquad M_{W^\prime} > (g_R/g_L) \times 620\ \textrm{GeV} .
\label{Bound_MWR}
\eeq
where $\zeta_g \equiv g_R^2 M_W^2 / g_L^2 M_{W^\prime}^2$.
In general, $\zeta_g \geq \xi_g \equiv (g_R/g_L)\xi$ for ordinary Higgs representations \cite{Langacker89,Nam02}.

  As well as $M_{W'}$ and $\xi_g$, we have additional NP parameters
such as $\theta_R$ and $\alpha_i$ in the quark sector as shown in Eq. (\ref{VR}) in the general LRM.
Among those new parameters, $\theta_R$ and $\alpha_2$ can be constrained
by the disagreement emerged between the determination of $|V_{ub}|$ from inclusive and exclusive B decays.
$|V_{ub}|$ determined in exclusive $B$ decays is related to $|V_{ub}^L|$ in the LRM as
\beq
|V_{ub}|_{excl} = |V^L_{ub}||1+\xi_u| \simeq |V_{ub}|_{incl}|1+\xi_u|  ,
\eeq
where $\xi_q \equiv \xi (g_RV^R_{qb})/(g_LV^L_{qb})$ and $q=u,c$ \cite{Nam08}.
From the mismatch between the values of $|V_{ub}|$ extracted from inclusive and exclusive $B$ decays,
we roughly obtain the following $2\sigma$ bound:
\beq \label{Bound_Vub}
-1.55 < \xi_g s_R\cos(\alpha_2 + \gamma) \times 10^3 < 0.41 ,
\eeq
where $\gamma = 68^\circ$.

  In Eq. (\ref{VR}), $\alpha_3$ and $\alpha_4$ are constrained by $B_s^0-\bar{B_s^0}$ mixing measurements.
The dispersive part of the $B_s^0-\bar{B_s^0}$ mixing matrix element in the LRM can be written as
\beq \label{massmixing}
M_{12}^s = M_{12}^{SM} + M_{12}^{LR} = M_{12}^{SM}\left( 1 + r^s_{LR} \right) ,
\eeq
where
\beq
r^s_{LR} \equiv \frac{M_{12}^{LR}}{M_{12}^{SM}}
  = \frac{\langle \bar{B_s^0}|H_{eff}^{LR}|B_s^0 \rangle}
         {\langle \bar{B_s^0}|H_{eff}^{SM}|B_s^0 \rangle} ,
\eeq
and the explicit form of the effective Hamiltonians $H_{eff}^{SM}$ and $H_{eff}^{LR}$
describing the $\Delta B = 2$ transition in the LRM can be found in Refs. \cite{Lee12, Nam02}.
Following the factorization methods used in Ref. \cite{Lee12} with the given form of $V^R$ in Eq. (\ref{VR}),
we obtain the right-handed current contributions to  $B_s^0-\bar{B_s^0}$ mixing as
\beqr \label{rLRs}
r^s_{LR} &\approx&  162 \biggl( \frac{ 1 - 5.03\zeta_g - (0.490 - 1.96\zeta_g )\ln(1/\zeta_g) }
         {1 - 10.2\zeta_g + 30.1\zeta_g^2}\biggr) \zeta_g s_Rc_Re^{-i(\alpha_3 - \alpha_4)} \
         + \ 1.70\xi_g s_R e^{-i\alpha_3}  .
\eeqr
The deviation of the present experimental data from the SM predictions on $B_s$ meson mixing gives
the following $2\sigma$ bound \cite{Lenz12}:
\beq \label{Bound_mixing}
0.86 < |1+r^s_{LR}| < 1.22 ,
\eeq
and we will use this bound together with those in Eqs. (\ref{Bound_MWR}) and (\ref{Bound_Vub})
for our numerical analysis in Sec V.

\section{Effective Hamiltonian}

 In order to include QCD effects systematically, we start from the following low-energy effective Hamiltonian
describing $\Delta B = 1$ and $\Delta S = 1$ transition as done similarly in Ref. \cite{Nam03}:
\beq
\mathcal{H}_{eff} = \frac{G_F}{\sqrt{2}} \left[
  \sum_{\substack{i=1,2,11,12 \\ q=u,c}}\lambda_q^{LL} C_i^q O_i^q  - \lambda_t^{LL}
  \left(\sum_{i=3}^{10} C_i O_i  + C_7^\gamma O_7^\gamma + C_8^G O_8^G\right) \right]
  + (C_i O_i \rightarrow C'_i O'_i) ,
\eeq
where $\lambda_q^{AB}\equiv V^{A\ast}_{qs}V^B_{qb}$,
$O_{1,2}$ are the standard current-current operators, $O_3 - O_{10}$ are
the standard penguin operators, and $O_7^\gamma$ and $O_8^G$ are the standard photonic and
gluonic magnetic operators, respectively, which can be found in Ref. \cite{Buchalla96}.
In addition to those SM operators, in the LRM, the operator basis is doubled by $O'_i$ which
are the chiral conjugates of $O_i$.  Also new operators $O_{11,12}$ and $O'_{11,12}$
arise with mixed chiral structure of $O_{1,2}$ and $O^\prime_{1,2}$ as follows:
\beqr
O^q_{11} = \left( \bar s_\alpha q_\beta \right)_{\rm V-A}
  \left( \bar q_\beta b_\alpha \right)_{\rm V+A}, &\qquad&
O^q_{12} = \left( \bar s_\alpha q_\alpha \right)_{\rm V-A}
  \left( \bar q_\beta b_\beta \right)_{\rm V+A},
\eeqr
where $(V\pm A)$ refers to the Lorentz structure $\gamma_\mu(1\pm\gamma_5)$.
These new operators may play an important role in tree-level dominated $b$ decays in the general LRM.

 The low-energy effects of the full theory at an arbitrary low-energy scale $\mu$ can then be described by
the linear combination of the given operators and the corresponding Wilson coefficients (WCs) $C_i(\mu)$.
In order to calculate $C_i(\mu)$, we first calculate them at $\mu = M_W$ scale.
After performing a straightforward matching computation, we find the WCs
including the electromagnetic penguin contributions at $W$ scale neglecting the $u$-quark mass:
\beqr \label{CoefMW}
C^q_2(M_W) &=& 1 , \quad
C^{q\prime}_2(M_W)\ =\ \zeta_g\lambda^{RR}_q/\lambda^{LL}_q \ (q=u,c), \cr
C_3(M_W) &=& \frac{\aem}{6\pi} \frac{1}{\sin^2\theta_W} \left[ 2 B(x_t) + C(x_t) \right] ,  \cr
C_7(M_W) &=& \frac{\aem}{6\pi} \left[ 4 C(x_t) + D(x_t)\right] ,\cr
C_9(M_W) &=& \frac{\aem}{6\pi} \left[ 4 C(x_t) + D(x_t) +
  \frac{1}{\sin^2\theta_W} (10 B(x_t) - 4 C(x_t)) \right] , \\
C_7^\gamma(M_W) &=& F(x_t) + \frac{m_t}{m_b}A^{tb}\widetilde{F}(x_t) , \quad
C_7^{\gamma\prime}(M_W)\ =\ \frac{m_t}{m_b}A^{ts\ast}\widetilde{F}(x_t) , \cr
C_8^G(M_W) &=& G(x_t) + \frac{m_t}{m_b}A^{tb}\widetilde{G}(x_t) , \quad
C_8^{G\prime}(M_W)\ =\ \frac{m_t}{m_b}A^{ts\ast}\widetilde{G}(x_t) , \cr
C^u_{12}(M_W) &=& A^{ub}, \quad C^{u\prime}_{12}(M_W)\ =\ A^{us\ast} , \nnb
\eeqr
where
\beq \label{CKMratio}
\qquad x_U = \frac{m_U^2}{M_W^2}\ (U=u,c,t), \qquad
A^{UD} = \xi_g \frac{V^R_{UD}}{V^L_{UD}}e^{i\alpha_\circ} \ (D=b,s),
\eeq
and $\alpha_\circ$ is a CP\ phase residing in the vacuum expectation values, which can be absorbed
in $\alpha_i$ in Eq. (\ref{VR}) by redefining $\alpha_i + \alpha_\circ \rightarrow \alpha_i$.
All other coefficients are negligible or vanish.  In Eq. (\ref{CoefMW}),
the explicit forms of the functions $B(x_t)$, $C(x_t)$, and $D(x_t)$ can be found
in Refs. \cite{Buchalla96, Buchalla90},
and $F(x_t)$, $\widetilde{F}(x_t)$, $G(x_t)$, and $\widetilde{G}(x_t)$ are given in Ref. \cite{Cho94}.
In the above magnetic coefficients, the terms proportional to $\xi_g$ and $\zeta_g$ are neglected
except the contribution coming from the virtual $t$ quark which gives $m_t/m_b$ enhancement.
Also the term proportional to $\zeta_g$ in the coefficient $C'_2$ is not neglected
because $\zeta_g \geq \xi_g$ and there is possible enhancement by the ratio of
CKM\ angles ($\lambda^{RR}_q/\lambda^{LL}_q$) in the nonmanifest LRM.
Note that the new coefficient $C^{u(\prime)}_{12}(M_W)$ can be important in some $b\to s$ transitions
because the $\xi_g$ suppression can be offset by the ratio $V^R_{uD}/V^L_{uD}$
in Eq. (\ref{CKMratio}).

  The coefficients $C_i(\mu)$ at the scale $\mu$ below $m_b$ can be obtained by evolving the coefficients
$C_i(M_W)$ with the $28\times 28$ anomalous dimension matrix applying the usual renormalization
group procedure in the following way:
\beq
\vec{C}(\mu) = U_4(\mu,m_b)M(m_b)U_5(m_b,M_W)\vec{C}(M_W) ,
\eeq
where $U_f$ is the evolution matrix for $f$ active flavors and $M(m)$ gives the matching corrections
between $\vec{C}_{f-1}(m)$ and $\vec{C}_f(m)$.
In the leading logarithmic (LL) approximation $M=1$ and the evolution matrix $U(m_1,m_2)$ is given by
\beq \label{RGevol}
U(m_1,m_2) = V\left[\left(\frac{\alpha_s(m_2)}{\alpha_s(m_1)}\right)^{\vec{\gamma}/2\beta_0}
 -\frac{\alpha}{2\beta_0}K(m_1,m_2)\right]V^{-1},
\eeq
where $V$ diagonalizes the transposed of the anomalous dimension matrix $\gamma_s$
and $\vec{\gamma}$ is the vector containing
the eigenvalues of $\gamma_s^{T}$. In the right-hand side of Eq. (\ref{RGevol}),
the first term represents the pure QCD\ evolution and the second term describes
the additional evolution in the presence of the electromagnetic interaction.
The leading order formula for the matrix $K(m_1,m_2)$ can be found in Refs. \cite{Buchalla96,Buchalla90}.
 Since the strong interaction preserves chirality, the $28\times 28$ anomalous
dimensional matrix decomposes into two identical $14\times 14$ blocks.  The SM\ $12\times 12$
submatrix describing the mixing among $O_1 - O_{10}$, $O_7^\gamma$, and $O_8^G$ can be found
in Ref. \cite{Ciuchini93}, and the explicit form of the remaining $4\times 4$ matrix describing
the mixing among $O_{11,12}$, $O_7^\gamma$, and $O_8^G$, which partially overlaps with the
SM\ $12\times 12$ submatrix, can be found in Ref. \cite{Cho94}.

In this paper, unlike the previous analysis in Ref. \cite{Nam03},
we set the scale of weak WCs at $\mu =$ 1.5 GeV
to use the PQCD results for the hadronic matrix elements \cite{Keum01}. For 4 flavors,
we have the following numerical values of $C_i$(1.5 GeV) in LL precision using
the standard quark masses:\footnote{Although QCD\ correction
factors in $C^\prime_{1,2}$ are different from those in $C_{1,2}$ in general,
we use an approximation $\alpha_s(M_{W^\prime})\simeq \alpha_s(M_W)$ for simplicity,
which will not change our result.}
\bdpm
C_1^q = -0.453 , \qquad C_1^{q\prime} = C_1^q\zeta_g\lambda^{RR}_q/\lambda^{LL}_q ,
\edpm
\bdpm
C_2^q = 1.231 , \qquad C_2^{q\prime} = C_2^q\zeta_g\lambda^{RR}_q/\lambda^{LL}_q ,
\edpm
\bdpm
C_3 = 0.024 , \quad C_4 = -0.046 , \quad C_5 = 0.012 , \quad C_6 = -0.066 ,
\edpm
\beq \label{Coef-num}
C_7 = 0.014\alpha , \quad C_8 = 0.069\alpha , \quad C_9 = -1.436\alpha ,\quad C_{10}= 0.503\alpha ,
\eeq
\bdpm
C_7^\gamma = -0.389 - 17.86A^{tb} , \qquad C_7^{\gamma\prime} = -17.86A^{ts\ast} ,
\edpm
\bdpm
C_8^G = -0.177 - 7.858A^{tb} , \qquad C_8^{G\prime} = -7.858A^{ts\ast}.
\edpm
\bdpm
C_{11}^{u} = 0.641A^{ub} , \quad C_{12}^{u} = 0.879A^{ub} , \quad
C_{11}^{u\prime} = 0.641A^{us\ast} , \quad C_{12}^{u\prime} = 0.879A^{us\ast} ,
\edpm
where subdominant NP terms are neglected.  Note that $C'_3 - C'_{10}$ are negligible comparing to $C_7^{\gamma\prime}$ and $C_8^{G\prime}$
whereas $C'_{1,2}$ and $C^{(\prime)}_{11,12}$ are not.  $C^{(\prime)}_{1,2}$ and $C^{(\prime)}_{11,12}$ can be
important especially to the tree-dominated $B$ decays.

\section{$B \to K \pi$ decay amplitudes }

 Following the procedure of Ref. \cite{Nam03} of including the penguin-type diagrams of the current-current operators
$O_{1,2}$ and the tree-level diagrams associated with the magnetic operators $O_7^\gamma$ and $O_8^G$,
the one-loop matrix elements of $\mathcal{H}_{eff}$ can be written
in terms of the tree-level matrix elements of the effective operators:
\beq
<sq\bar{q}|\mathcal{H}_{eff}|B> = -\frac{G_F}{\sqrt{2}}\lambda^{LL}_t
\sum_{i=1}^{12}C_i^{eff}<sq\bar{q}|O_i|B>^{tree} +\ (C_i^{eff} O_i \rightarrow C^{eff\prime}_i O'_i),
\eeq
with the effective WCs
\beqr
C_i^{eff(\prime)} = C_i^{(\prime)} && (i=1,2,8,10,11,12) , \cr
C_3^{eff(\prime)} = C_3^{(\prime)} - \frac{1}{N_c}C_g^{(\prime)} ,
 && C_4^{eff(\prime)} = C_4^{(\prime)} + C_g^{(\prime)} , \cr
C_5^{eff(\prime)} = C_5^{(\prime)} - \frac{1}{N_c}C_g^{(\prime)} ,
 && C_6^{eff(\prime)} = C_6^{(\prime)} + C_g^{(\prime)} , \\
C_7^{eff(\prime)} = C_7^{(\prime)} + C_\gamma^{(\prime)} ,
 && C_9^{eff(\prime)} = C_9^{(\prime)} + C_\gamma^{(\prime)} , \nnb
\eeqr
where
\beqr
C_g^{(\prime)} &=& -\frac{\alpha_s}{8\pi}\left[\frac{1}{\lambda_t^{LL}}\sum_{q=u,c}\lambda_q^{LL} C_2^{q(\prime)}
 \mathcal{I}(m_q,k,m_b) + 2C_8^{G(\prime)}\frac{m_b^2}{k^2} \right] ,\\
C_\gamma^{(\prime)} &=& -\frac{\alpha}{3\pi}\left[\frac{1}{\lambda_t^{LL}}\sum_{q=u,c}\lambda_q^{LL}(C_1^{q(\prime)}
 +\frac{1}{N_c}C_2^{q(\prime)})\mathcal{I}(m_q,k,m_b) + C_7^{\gamma(\prime)} \frac{m_b^2}{k^2} \right] ,\nnb
\eeqr
and
\beq
\mathcal{I}(m,k,\mu) = 4\int_0^1 dx x(1-x)\ln \Big[\frac{m^2 - k^2x(1-x)}{\mu^2}\Big] .
\eeq
and where $k$ is the momentum transferred by the photon or the gluon to the ($q,\bar{q}$) pair.
Here $k^2$ is expected to be typically in the range $m_b^2/4\leq k^2\leq m_b^2/2$ \cite{Desh90},
and we will use $k^2=m_b^2/2$ for our numerical analysis.
The expression of the decay amplitudes can be further simplified by combining the effective WCs
in the following way:
\beq
\ a_{2i-1}^{(\prime)} = C^{eff(\prime)}_{2i-1}+\frac{1}{N_c}C^{eff(\prime)}_{2i}, \quad
a_{2i}^{(\prime)} = C^{eff(\prime)}_{2i}+\frac{1}{N_c}C^{eff(\prime)}_{2i-1} \ (i=1,2,3),
\eeq
where the factor $1/N_c$ originates from fierzing the operators $O_i^{(\prime)}$
after adopting the factorization assumption, and $N_c$ is simply equal to the number of colors
in the naive factorization approximation based on the vacuum-insertion method \cite{Gaillard74}.
Also, in the PQCD approach, $N_c \approx 3$ as well because of the cancellation between
the nonfactorizable contributions.

 The matrix amplitudes for $B \to K \pi$ decays can then be written in terms of the effective WCs as
\beqr \label{Amplitudes}
\mathcal{A}(\bar{B}^0\rightarrow \pi^0 \bar{K}^0) &=& \frac{G_F}{2}
 \Bigg\{ \left[\lambda_u^{LL}(a_1 + \rho^\pi_u a_{11})+\frac{3}{2}\lambda_t^{LL}(a_7-a_9)\right]X^{(BK,\pi)}  \cr
&& + \lambda_t^{LL}\left[a_4-\frac{1}{2}a_{10}+2\rho^K_s\left(a_6-\frac{1}{2}a_8\right)\right]X^{(B\pi,K)} \cr
&& + \lambda_t^{LL}\left[a_4-\frac{1}{2}a_{10}+2\rho^B_s\left(a_6-\frac{1}{2}a_8\right)\right]X^{(B,\pi K)}\Bigg\} \cr
&& +\, (a_i \to -a_i^\prime), \\
\mathcal{A}(\bar{B}^0\rightarrow \pi^+ K^-) &=& \frac{G_F}{\sqrt{2}}
 \Bigg\{ \left[\lambda_u^{LL}(a_2 + a_{12})-\lambda_t^{LL}\Big(a_4+a_{10}+2\rho^K_s(a_6+a_8)\Big)\right]X^{(B\pi,K)} \cr
&& - \lambda_t^{LL}\left[a_4-\frac{1}{2}a_{10}+2\rho^B_s\left(a_6-\frac{1}{2}a_8\right)\right] X^{(B,\pi K)}\Bigg\} \cr
&&\, + (a_i \to -a_i^\prime), \\
\mathcal{A}(B^-\rightarrow \pi^0 K^-) &=& \frac{G_F}{2}
 \Bigg\{ \left[\lambda_u^{LL}(a_1 + \rho^\pi_u a_{11})+\frac{3}{2}\lambda_t^{LL}(a_7-a_9)\right]X^{(BK,\pi)}  \cr
&& + \left[ \lambda_u^{LL}(a_2 + a_{12}) -\lambda_t^{LL}\Big(a_4+a_{10}+2\rho^K_s(a_6+a_8)\Big)\right]X^{(B\pi,K)} \cr
&& + \left[ \lambda_u^{LL}(a_2 - a_{12}) -\lambda_t^{LL}\Big(a_4+a_{10}+2\rho^B_s(a_6+a_8)\Big)\right]
 X^{(B,\pi K)}\Bigg\} \cr
&&\, + (a_i \to -a_i^\prime), \\
\mathcal{A}(B^-\rightarrow \pi^- \bar{K}^0) &=& \frac{G_F}{\sqrt{2}}
 \Bigg\{ -\lambda_t^{LL}\left[a_4-\frac{1}{2}a_{10}
 +2\rho^K_s\left(a_6-\frac{1}{2}a_8\right)\right]X^{(B\pi,K)} \cr
&& +\left[\lambda_u^{LL} (a_2 - a_{12}) - \lambda_t^{LL}\Big(a_4+a_{10}+2\rho^B_s\left(a_6+a_8\right)\Big)\right] X^{(B,\pi K)}\Bigg\} \cr
&&\, + (a_i \to -a_i^\prime),
\eeqr
where
\beqr \label{formfactor}
X^{(BK,\pi)} &=& -\sqrt{2}<\pi^0|\bar{u}\gamma^\mu\gamma_5 u|0><\bar{K}^0|\bar{s}\gamma_\mu b|\bar{B}^0> \cr
  &=& i f_\pi F_0^{B\to K}(m_\pi^2)(m_B^2-m_K^2) , \cr
X^{(B\pi,K)} &=& +\sqrt{2}<\bar{K}^0|\bar{s}\gamma^\mu\gamma_5 d|0><\pi^0|\bar{d}\gamma_\mu b|\bar{B}^0> \cr
  &=& i f_K F_0^{B\to \pi}(m_K^2)(m_B^2-m_\pi^2) , \cr
X^{(B,\pi K)} &=& +\sqrt{2}<\pi^0\bar{K}^0|\bar{s}\gamma^\mu d|0><0|\bar{d}\gamma_\mu\gamma_5 b|\bar{B}^0> \cr
  &=& i f_B F_0^{\pi K}(m_B^2)(m_K^2-m_\pi^2) ,
\eeqr
and where
\beq
\rho^H_q \equiv \frac{m_H^2}{m_bm_q} \ (H=\pi,K,B ,\ q=u,s).
\eeq
Note from Eq. (\ref{formfactor}) that the form factors $F_0^{B\to K}$ and $F_0^{B\to \pi}$ can be determined by
relevant semileptonic $B$ decays, but $F_0^{\pi K}$ is not.
Because of the significant hadronic uncertainties in the factorization approximation of the matrix amplitudes,
it is very difficult to separately determine the size of NP contributions.
Therefore, in this paper, instead of performing a complete analysis by varying all relevant independent NP parameters
($\zeta_g, \xi_g, \theta_R, \alpha_{2,3,4}$) in this model,
we fix $\xi_g$ and $\alpha_{1,3,4}$ for simple illustration of NP effects.
Also, for numerical analysis, we use the following values of form factors obtained from PQCD calculation \cite{Chen06}:
\beq
F_0^{B\to K}(m_\pi^2) = 0.37 , \quad F_0^{B\to \pi}(m_K^2) = 0.24 , \quad
F_0^{\pi K}(m_B^2) = (0.39+8.16i)10^{-4} .
\eeq
The form factor $F_0^{\pi K}(m_B^2)$ which originates from annihilation contributions is complex 
due to the final state quark interactions, so that it could be important in CP observables.  
Also, due to the enhancement factor $\rho^B_s$ proportional to $m^2_B$, 
the annihilation contributions are not negligible.

\section{Results}

\begin{figure}[!hbt]
\centering%
  \subfigure[$\alpha_1=\alpha_3=\pi$]{\label{Phase1} %
    \includegraphics[height=7cm]{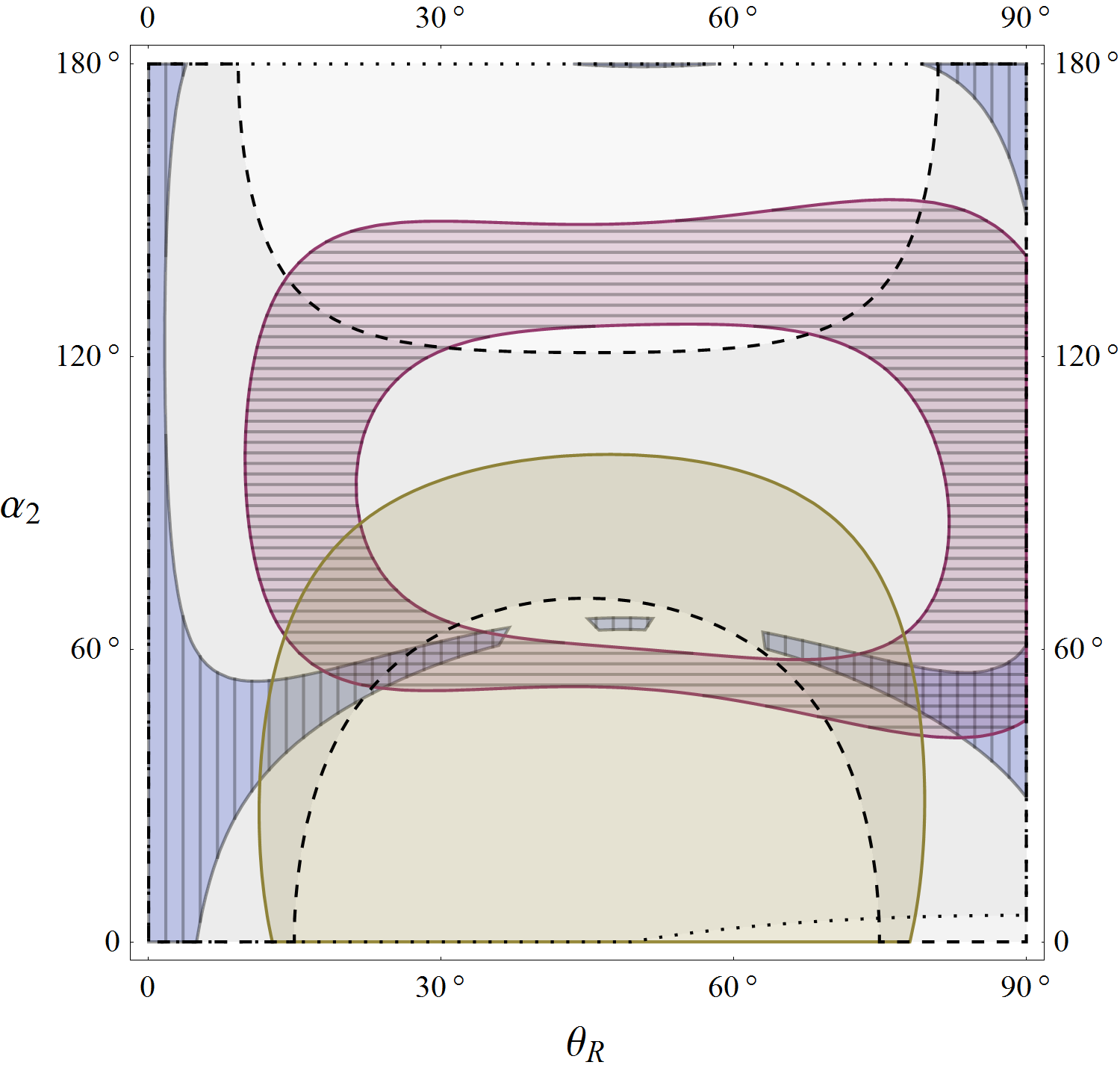}} \qquad
  \subfigure[$\alpha_1=\pi, \alpha_3=-\pi/2$]{\label{Phase2} %
    \includegraphics[height=7cm]{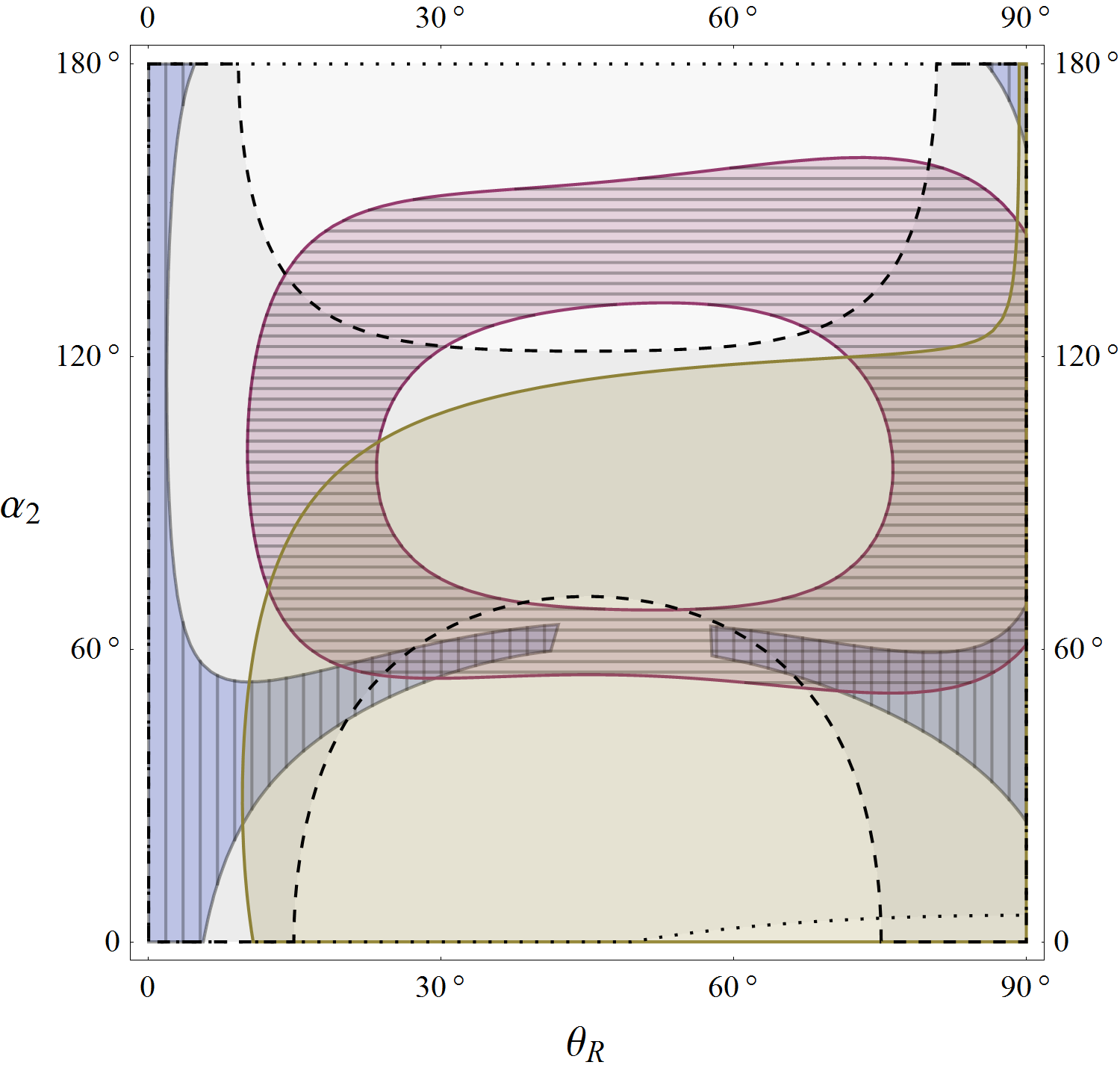}}
\caption{Allowed regions for $\alpha_2$ and $\theta_R$ at 2 $\sigma$ level for $M_{W'} = (g_R/g_L)\times 1.5$ TeV.
The blue(vertical mesh), red(horizontal mesh), and yellow(no mesh) regions are allowed by the current measurements
of $A_{CP}(B^0 \to K^\pm \pi^\mp)$, $A_{CP}(B^\pm \to K^\pm \pi^0)$, and $A_{CP}(B^\pm \to K^0 \pi^\pm)$, respectively.
The dotted and dashed lines indicate the bounds by the measurements of $V_{ub}$ and $B_s^0-\bar{B_s^0}$ mixing,
respectively.}
\label{Phase}
\end{figure}

\begin{figure}[!hbt]
\centering%
  \subfigure[$\alpha_1=\alpha_3=\pi$]{\label{Angle1} %
    \includegraphics[height=7cm]{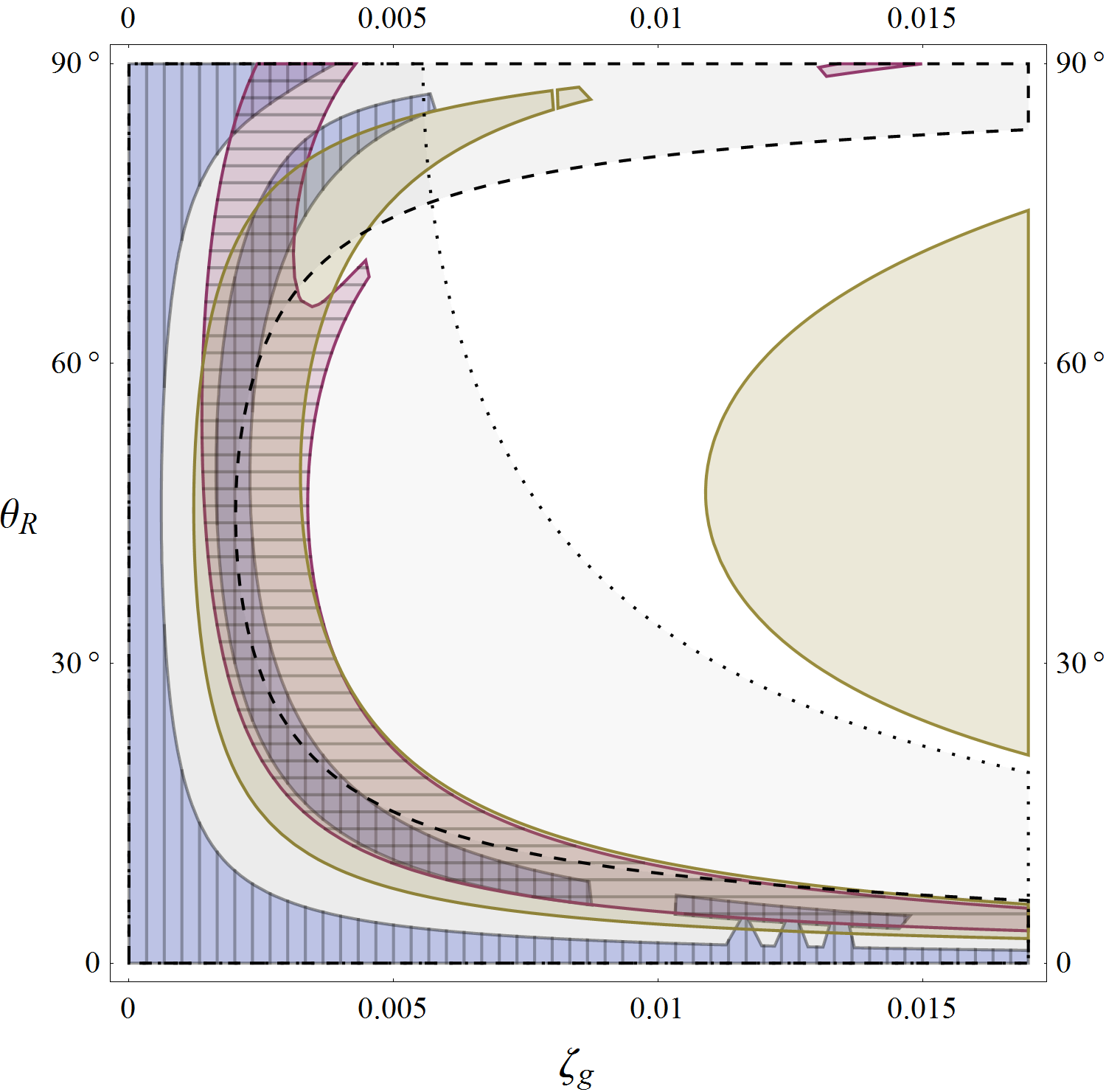}} \qquad
  \subfigure[$\alpha_1=\pi, \alpha_3=-\pi/2$]{\label{Angle2} %
    \includegraphics[height=7cm]{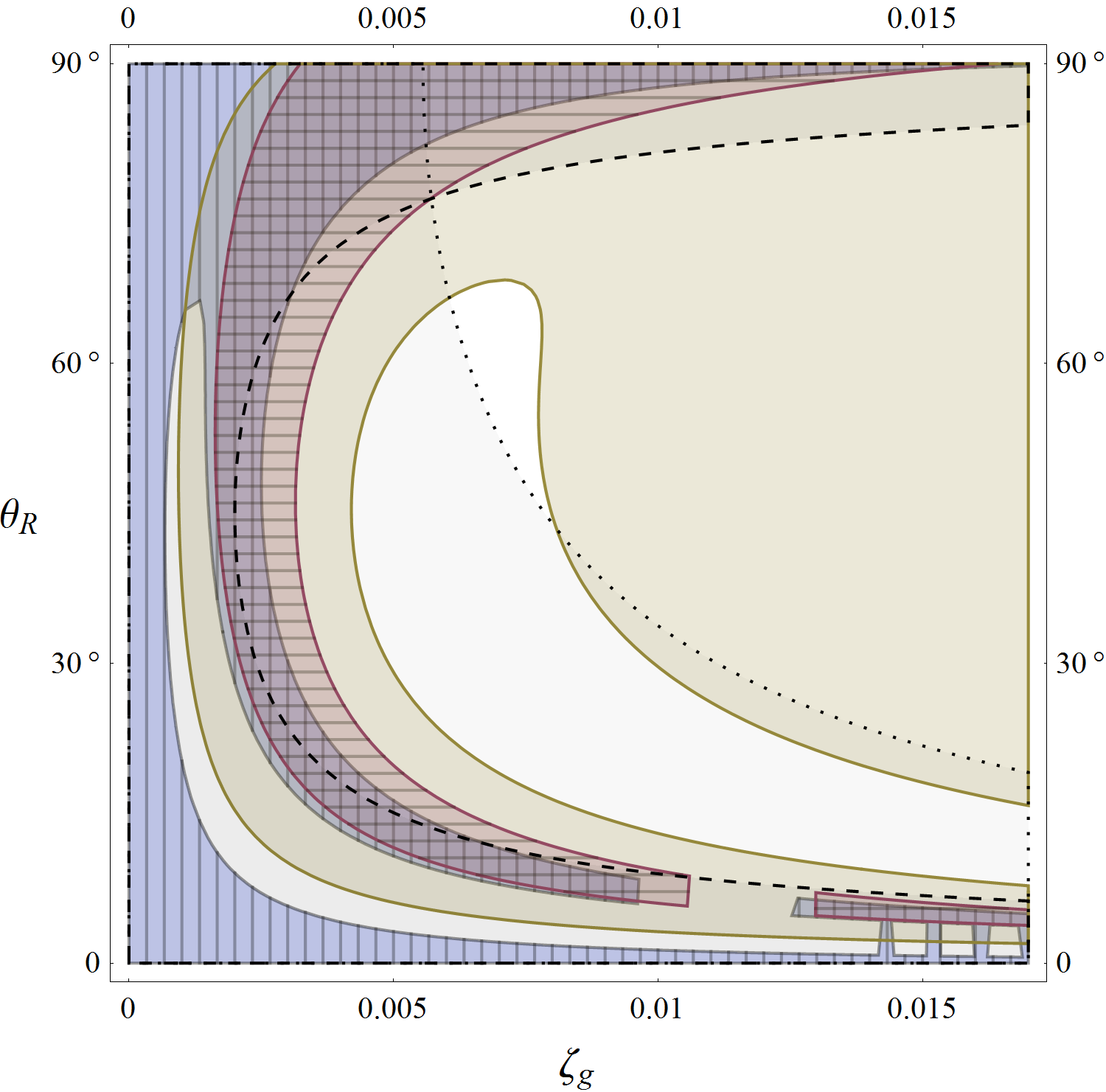}}
\caption{Allowed regions for $\theta_R$ and $\zeta_g$ at 2 $\sigma$ level.
The blue(vertical mesh), red(horizontal mesh), and yellow(no mesh) regions are allowed by the current measurements
of $A_{CP}(B^0 \to K^\pm \pi^\mp)$, $A_{CP}(B^\pm \to K^\pm \pi^0)$, and $A_{CP}(B^\pm \to K^0 \pi^\pm)$, respectively.
The dotted and dashed lines indicate the bounds by the measurements of $V_{ub}$ and $B_s^0-\bar{B_s^0}$ mixing,
respectively.}
\label{Angle}
\end{figure}

 For illustration of allowed new model parameter spaces,
we first fix the new gauge boson mass to be $M_{W'} = (g_R/g_L)\times 1.5$ TeV,
and plot the allowed region of $\alpha_2$ and $\theta_R$ at $2\sigma$ level
for $\alpha_1=\alpha_3=\pi$ in Fig. \ref{Phase1} and for $\alpha_1=\pi, \alpha_3=-\pi/2$ in Fig. \ref{Phase2}
using the present experimental bounds of the CP asymmetries in Eq. (\ref{ACP_exp}).
With the chosen NP parameters, the branching fraction of each decay mode in Eq. (\ref{Amplitudes})
agrees with the present experimental measurement as well.
In Fig. \ref{Phase}, the region (nearly all) above the dotted line is allowed
by the $V_{ub}$ bound given in Eq. (\ref{Bound_Vub}),
and the shaded area inside the dashed lines is allowed
by the $B_s^0-\bar{B_s^0}$ mixing bound in Eq. (\ref{Bound_mixing}).
From the overlapped allowed regions of the figures, one can see that the value of $\theta_R$ could be
either small or large, but nonzero value of $\alpha_2$ near $60^\circ$ is preferred in both cases similarly.

 In order to clearly see $W'$ mass dependence, we plot the allowed region of $\theta_R$ and $\zeta_g$ at $2\sigma$ level
for $\alpha_2=55^\circ$ in Fig. \ref{Angle},
taking into account the constraint from the muon decay measurements given in Eq. (\ref{Bound_MWR}).
In Fig. \ref{Angle}, similarly, the shaded regions left of the dotted and dashed lines are allowed
by the $V_{ub}$ bound given in Eq. (\ref{Bound_Vub})
and by the $B_s^0-\bar{B_s^0}$ mixing bound in Eq. (\ref{Bound_mixing}), respectively.
With the given parameter sets, we estimate the size of the right-handed current contributions 
responsible for the present measurements
of $A_{CP}(B^0 \to K^\pm \pi^\mp)$, $A_{CP}(B^\pm \to K^\pm \pi^0)$, and $A_{CP}(B^\pm \to K^0 \pi^\pm)$
as shown in Fig. \ref{Angle}, and obtain the lower bound of $\zeta_g$ approximately given as $\zeta_g \gtrsim 0.0015$
which corresponds to the upper bound of $W'$ mass $M_{W'} \lesssim (g_R/g_L)\times 2.1$ TeV.
We found that this mass bound could be somewhat higher for different values of $\alpha_2$,
but not drastically different.
It should also be noted that we scanned other sets of NP parameters,
and have no better results (no wider simultaneously allowed regions)
for different values of $\alpha_{1,3,4}$.

\section{Concluding Remarks}

In this paper, we studied the right-handed current contributions to the direct CP asymmetries in $B \to K \pi$ decays
taking into account all possible tree and penguin contributions including annihilation-type amplitudes 
by adopting PQCD approach  in the nonmanifest LRM.
Without imposing manifest or pseudomanifest left-right symmetry,
we parametrized $V^R$ as shown in Eq. (\ref{VR})
so that $W'$ mass is not strongly constrained by the current direct and indirect search results \cite{Langacker89,Buras11},
and showed that the CP asymmetries are sensitive to
the phases and angles in $V^R$ as well as to the mass of $W'$.
We considered the constraints from the global analysis of muon decay parameters, 
the determination of $|V_{ub}|$ in inclusive and exclusive $B$ decays,
and $B_s^0-\bar{B_s^0}$ mixing measurements.
With the given phases, one can see from the figures that
relatively large value of the mixing angle $\theta_R$ is preferred unless $W'$ mass is as light as a few hundred GeV.
This could also give simultaneous explanations to the large branching fractions of $B \to \tau \nu$ transitions
due to a large fraction $V^R_{ub}/V^L_{ub}$ \cite{Nam08},
and also to the large CP-violating like-sign dimuon charge asymmetry in semileptonic B decays \cite{Lee12}
due to a large CP-violating phase in $B_s$ mixing \cite{Buras11}.
Also, with the given parameter sets, Fig. \ref{Angle} shows that it is favorable that
the mass of $W'$ is lighter than around  $(g_R/g_L)\times 2.1$ TeV
in order to incorporate the current experimental measurements.
In this way, CP asymmetries in other nonleptonic B decays such as $B \to K \rho$ and $B \to K^\ast \pi$ can be estimated systematically and similarly, and all of these analysis of possible NP contributions can be tested
once future experimental progress can further improve the bounds.

\begin{acknowledgments}

S.-h. Nam thanks C.-H. Chen and H-n. Li for useful communications, and
C.-W. Chiang for collaboration at the beginning of this project.
The computation of this work was supported in part by the PLSI supercomputing resources of KISTI.

\end{acknowledgments}

\def\epjc#1#2#3 {Eur. Phys. J. C {\bf#1}, #2 (#3)}
\def\jhep#1#2#3 {JHEP {\bf#1}, #2 (#3)}
\def\npb#1#2#3 {Nucl. Phys. B {\bf#1}, #2 (#3)}
\def\mpla#1#2#3 {Mod. Phys. Lett. A {\bf#1}, #2 (#3)}
\def\plb#1#2#3 {Phys. Lett. B {\bf#1}, #2 (#3)}
\def\prd#1#2#3 {Phys. Rev. D {\bf#1}, #2 (#3)}
\def\prl#1#2#3 {Phys. Rev. Lett. {\bf#1}, #2 (#3)}
\def\ptp#1#2#3 {Prog. Theor. Phys. {\bf#1}, #2 (#3)}
\def\rmp#1#2#3 {Rev. Mod. Phys. {\bf#1}, #2 (#3)}
\def\zpc#1#2#3 {Z. Phys. C {\bf#1}, #2 (#3)}
\def\ibid#1#2#3 {{\it ibid.} {\bf#1}, #2 (#3)}
\def\none#1#2#3 {{\bf#1}, #2 (#3)}

\end{document}